\documentclass[aps,prd,amsmath,amssymb,twocolumn,superscriptaddress,preprintnumbers,nofootinbib]{revtex4-1}
\pdfoutput=1
\usepackage{graphicx}
\usepackage{amsthm}
\usepackage{amsmath}
\usepackage{graphicx,subfigure}
\usepackage{dcolumn}
\usepackage{bm}
\usepackage{slashed}
\usepackage{color}

\newcommand{\be}{\begin{eqnarray}}
\newcommand{\ee}{\end{eqnarray}}
\newcommand{\bea}{\begin{eqnarray}}
\newcommand{\eea}{\end{eqnarray}}

\newcommand{\GeV}{{~\rm GeV}}
\newcommand{\gev}{{~\rm GeV}}
\newcommand{\TeV}{{~\rm TeV}}
\newcommand{\tev}{{~\rm TeV}}

\newcommand{\vev}[1]{{\langle #1 \rangle}}

\newcommand{\mX}{m_{_\chi} }

\newcommand{\thetaX}{\theta_{_\chi}}
\newcommand{\thetaXS}{\theta_{_\chi}}
\newcommand{\thetaXA}{\theta_{_\chi}}

\newcommand{\mZ}{m_{\rm Z}}
\newcommand{\mW}{m_{\rm W}}
\newcommand{\LamR}{\Lambda_{_R}}
\newcommand{\LamRt}{\tilde{\Lambda}_{_R}}

\newcommand{\muN}{\mu_{_N}}
\newcommand{\muX}{\mu_{_\chi}}

\newcommand{\ZZ}{{\rm Z}}
\newcommand{\Wpm}{\rm W^\pm}
\newcommand{\Wmp}{\rm W^\mp}

\newcommand{\cW}{{\rm c_{_W}}}
\newcommand{\sW}{{\rm s_{_W}}}

\newcommand{\SUWeak}{{\rm SU_{_W}(2)}}
\newcommand{\medf}{\psi}
\newcommand{\meds}{\varphi}
\newcommand{\mmedf}{M_f}
\newcommand{\mmeds}{M_s}
\newcommand{\loopint}[2]{\int \frac{d^{#1}#2}{(2\pi)^{#1}}}

\begin{document}

\title{UV Completions of Magnetic Inelastic Dark Matter and RayDM for the Fermi Line(s)}
\author{Neal Weiner}
\email{neal.weiner@nyu.edu}
\affiliation{Center for Cosmology and Particle Physics, Department of Physics, New York University, New York, NY 10003}
\author{Itay Yavin}
\email{iyavin@perimeterinstitute.ca}
\affiliation{Department of Physics \& Astronomy, McMaster University 1280 Main St. W. Hamilton, Ontario, Canada, L8S 4L8}
\affiliation{Perimeter Institute for Theoretical Physics 31 Caroline St. N, Waterloo, Ontario, Canada N2L 2Y5.}


\begin{abstract}
Models that seek to produce a line at $\sim$130 GeV as possibly present in the Fermi data face a number of phenomenological hurdles, not the least of which is achieving the high cross section into $\gamma \gamma$ required. A simple explanation is a fermionic dark matter particle that couples to photons through loops of charged messengers. We study the size of the dimension 5 dipole (for a pseudo-Dirac state) and dimension 7 Rayleigh operators in such a model, including all higher order corrections in $1/M_{mess}$. Such corrections tend to enhance the annihilation rates beyond the naive effective operators. We find that while freezeout is generally dominated by the dipole, the present day gamma-ray signatures are dominated by the Rayleigh operator, except at the most strongly coupled points, motivating a hybrid approach. With this, the Magnetic inelastic Dark Matter scenario provides a successful explanation of the lines at only moderately strong coupling. We also consider the pure Majorana WIMP, where both freezeout and the Fermi lines can be explained, but only at very strong coupling with light ($\sim 200 \-- 300 \gev$) messengers. In both cases there is no conflict with non-observation of continuum photons.
\end{abstract}

\maketitle

\section{Introduction}
One of the most striking signals for indirect detection of dark matter is a monoenergetic gamma ray line. Such a feature, in different locations, with the same energy has no known astrophysical background. While by its very nature dark matter must only couple very weakly to photons if at all, even a small signal could be very convincing. In light of this, the recent claims of a line from the galactic center at approximately 130 GeV \cite{Bringmann:2012vr,Weniger:2012tx,Tempel:2012ey,Su:2012ft} (and possibly a second near 111 GeV\cite{Su:2012ft}) are extremely exciting. Subsequent and statistically weaker claims to have seen a consistent signal in galaxy clusters \cite{Hektor:2012kc} and unassociated point sources \cite{Su:2012zg} (although see \cite{Hooper:2012qc,Hektor:2012jc}), while individually perhaps unpersuasive, make the possibility even more intriguing.

There are possible concerns, of course. The location of the peak in the GC seems not precisely central \cite{Su:2012ft} (although see \cite{Yang:2012he}), contrary to naive expectations. However, some numerical simulations that accounts for both dark matter as well as baryons have found possible preliminary evidence for such a transient feature \cite{Kuhlen:2012he}. No assoicated continuum emission has been detected from the galactic center \cite{Buchmuller:2012rc,Cohen:2012me,Cholis:2012fb}. Diffuse limits provide tensions, but no clear exclusions \cite{Huang:2012yf}. A strange peak at the same energy has been found in a subset of the Earth limb photons \cite{finkbeineridmtalk}, raising some concern that a bizarre instrumental effect could be giving rise to this. At the same time, no compelling explanation has been offered that would yield a sharp peak in the GC and not in e.g., the disk as well.

Regardless of these issues, it is worth understanding what models could explain these data. A number have been put forward (e.g., \cite{Cline:2012nw, Rajaraman:2012db,Buckley:2012ws,Das:2012ys,Weiner:2012cb,Heo:2012dk,Park:2012xq,Tulin:2012uq,Cline:2012bz,Bai:2012qy,Bringmann:2012mx,Bergstrom:2012mx}). An effective theory approach was taken in \cite{Weiner:2012cb}. There, it was noted that the interactions  of a neutral fermionic field with the electroweak gauge-bosons of the SM are constrained by gauge-invariance to the magnetic dipole interaction
\be
\label{eqn:MiDMLagrangian}
\mathcal{L}_{\rm MiDM} &=& \left(\frac{\mu_\chi}{2}\right)\bar \chi^* \sigma_{\mu\nu} B^{\mu\nu} \chi + c.c., \
\ee
and the Rayleigh operators,
\be
\mathcal{L}_{\rm RayDM} &=& \frac{1}{4\LamR^3}~ \bar{\chi}\chi \left( \cos\thetaXS B_{\mu\nu}B^{\mu\nu} + \sin\thetaXS{\rm Tr} W_{\mu\nu}W^{\mu\nu} \right) \\\nonumber  &+&\frac{i}{4\LamRt^{3}}~\bar{\chi}\gamma_5\chi \left( \cos\thetaXA B_{\mu\nu}\tilde{B}^{\mu\nu} + \sin\thetaXA{\rm Tr} W_{\mu\nu}\tilde{W}^{\mu\nu} \right).
\ee 
where in general $\LamR \ne \LamRt$ as we shall see below. Here $\chi$ and $\chi^*$ are distinct Weyl fermions which may be two independent Majorana fermions or part of a pseudo-Dirac pair. Ref.~\cite{Weiner:2012cb} showed that in order to obtain the annihilation rate associated with the $130\GeV$ line reported in Ref.~\cite{Bringmann:2012vr,Weniger:2012tx,Tempel:2012ey,Su:2012ft}  the necessary dipole strength in the case of MiDM corresponds to a scale of about $\muX^{-1} \sim 2\TeV$ whereas the necessary Rayleigh scale in the case of RayDM is in the range $\LamR \sim 500-600\GeV$\footnote{Monochromatic gammas from the dipole operator in MiDM was first discussed by \cite{Goodman:2010qn}. A more recent discussion where the early universe annihilation was suppressed by a large $\sim 20\gev$ splitting (and a somewhat larger dipole) was recently considered in \cite{Tulin:2012uq}.}. Naive dimensional analysis then implies that the physics that resolves these non-renormalizable operators is somewhat strongly interacting (a point also made in \cite{Buckley:2012ws,Weiner:2012cb,Cline:2012bz}).  In this letter we provide a concrete example of such UV physics with a simple renormalizable model that gives rise to a WIMP with both magnetic dipole and Rayleigh type interactions as featured in the MiDM and RayDM models. This UV completion helps to clarify the roles played by each of these interactions and their relative importance in different phases of the theory. 
 
\section{The Model}

In addition to the WIMP state $\chi$ which is a Dirac fermion, we consider a messenger state, a Dirac fermion $\medf$ and a charged scalar $\meds$, both of which are $\SUWeak$ doublets with hypercharge $Y=1/2$ and are heavier than the WIMP. They couple to the WIMP state through a Yukawa coupling which we denote by $\lambda$. The Lagrangian for this model is given by 
\be
\nonumber
\mathcal{L} &=& \bar{\chi}\left( i\slashed{\partial} - \mX\right)\chi - \tfrac{1}{2}\delta m~ \chi C\chi+ \bar{\medf}\left( i\slashed{D} - \mmedf\right)\medf 
\\
&+& \left(D^\mu\meds \right)^\dag D_\mu\meds - \mmeds^2 \meds^\dag \meds+ \lambda \bar{\medf}\chi \meds + h.c. 
\ee
where $D_\mu = \partial_\mu - i g W_\mu^a\tau^a - i \tfrac{1}{2} g' B_\mu$ is the covariant derivative associated with the $\SUWeak \times U_Y(1)$ gauge-bosons, $W_\mu^a$ and $B_\mu$, respectively, and $\tau^a$ are the $\SUWeak$ generators obeying tr$\left(\tau^a\tau^b\right) = \tfrac{1}{2} \delta^{ab}$ and related to the Pauli matrices through $\tau^a=\tfrac{1}{2} \sigma^a$. Aside from its Dirac mass, $\mX$, the WIMP states are split by a Majorana mass $\delta m$. 

When the mass of the WIMP is much lower than that of the messengers, its interactions with light fields such as the photon and weak vector-bosons can be described by an effective Lagrangian. Gauge invariance forces these interactions to appear as dimension 5, magnetic dipole operator as well as dimension 7, Rayleigh operators\footnote{After EWSB other, lower dimensional operators may appear involving the Higgs field, however those appear at higher loop order and are correspondingly much further suppressed.}.   Since the model above is a renormalizable interacting theory these operators can be computed in perturbation theory. However, because we will be dealing with scenarios where the new states are not much heavier than the dark matter, it is important to include $\mX/\mmedf$ corrections to these new operators (i.e., the form factors). In this letter we include all $\mX/\mmedf$ effects at 1-loop order when computing the non-relativistic cross-sections relevant for phenomenology.

\begin{figure}
\begin{center}
\includegraphics[width=0.23 \textwidth]{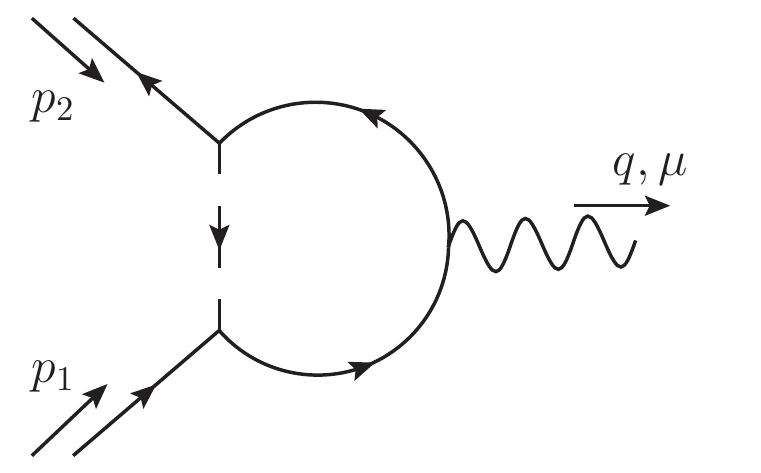}
\includegraphics[width=0.23 \textwidth]{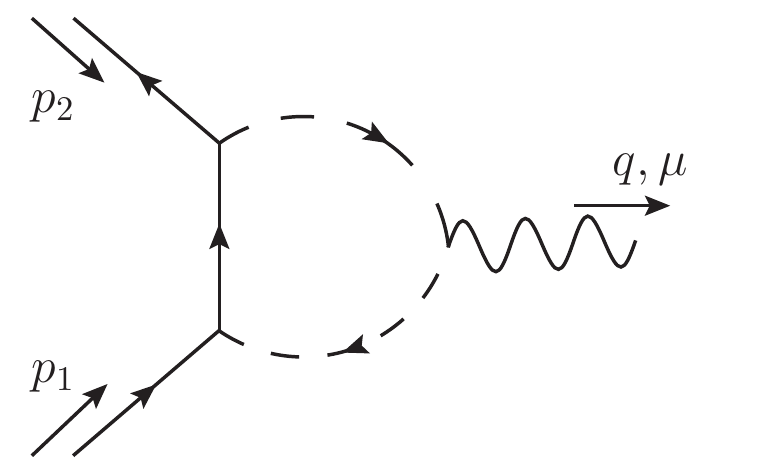}
\end{center}
\caption{Magnetic dipole operator generated at 1-loop.}
\label{fig:dipole_loop}
\end{figure}

We begin with the interactions of the WIMP with a single gauge-boson. These are generated through the diagram shown in Fig.~\ref{fig:dipole_loop}. Gauge-invariance forbids any coupling to the non-abelian $\SUWeak$ fields and the most general vertex coupling to hypercharge consistent with Lorentz invariance can be written as,
\be
\Gamma^\mu(q^2) = \gamma^\mu F_1(q^2) + i\left( \frac{\mu_\chi}{2}\right) \sigma^{\mu\nu}q_\nu  ~F_2(q^2)
\ee
where the form-factors $F_1(q^2)$ and $F_2(q^2)$ are given explicitly in the appendix\footnote{The $F_1(q^2)$ form-factor need not vanish as it is related to non-renormalizable terms of the form $\bar{\chi}\gamma^\mu\partial^\nu\chi B_{\mu\nu}$. Gauge-invariance only imposes the condition that $F_1(q^2)$ should approach zero as $q^2 \rightarrow 0$.}. The second part of this vertex corresponds to an effective dipole operator for the WIMP $\left(\frac{\muX}{2}\right)\bar{\chi} \sigma_{\mu\nu} B^{\mu\nu} \chi$ with the dipole strength being
\be
\label{eqn:UVforMiDM}
\mu_\chi = \frac{\lambda^2 g'}{32\pi^2\mmedf } 
\ee
where $g'$ is the hypercharge coupling constant, $q^2$ is the momentum carried by the gauge-boson. More explicitly, the coefficient of the  dipole operator  is multiplied by the hypercharge and by the size of the $\SUWeak$ representation of the messengers in the loop, which in our case gives a factor of unity. Similar comments apply to the coefficient of $F_1(q^2)$. To lowest order in an expansion in the messenger mass these form-factors are 
\be
F_1(q^2) &=& -\frac{\mu_\chi q^2}{6\mmedf}\left(\frac{2r^2 \left(3 r^2-3-\left(2+r^2\right) \log r^2\right)}{ \left(1-r^2\right)^2}\right) \\
F_2(q^2) &=& \frac{2 r^2 \left(r^2-1-\log r^2\right)}{\left(1-r^2\right)^2} 
\ee
where $r = \mmedf/\mmeds$. We include the effects of both $F_1$ and $F_2$ to all order in the messenger mass expansion in the cross-sections discussed below. 

\begin{figure}
\begin{center}
\includegraphics[width=0.5 \textwidth]{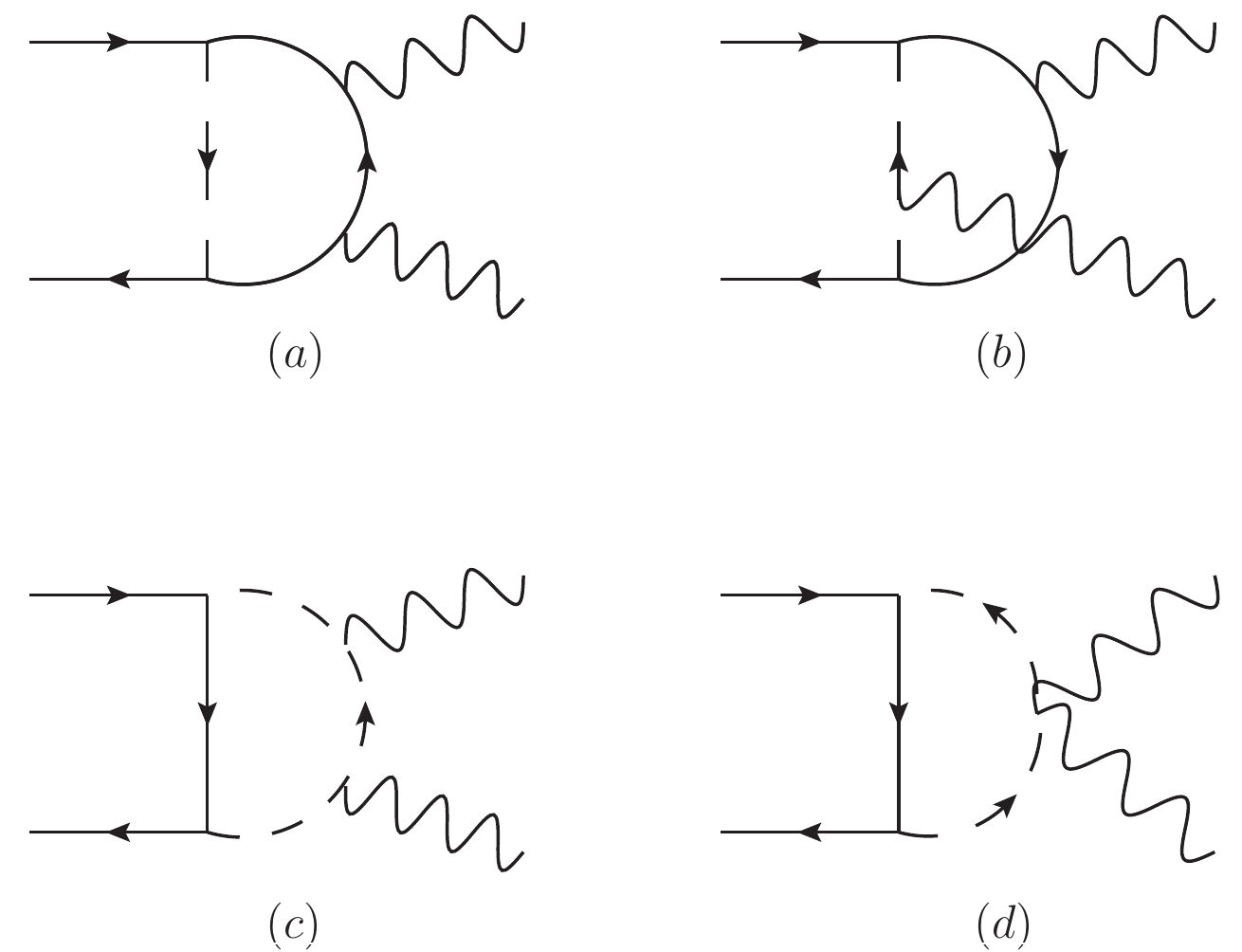}
\end{center}
\caption{The loop diagrams generating the RayDM operators at lowest order in perturbation theory. Diagrams (a), (b), and (c) represent two separate contributions where the external gauge-bosons are interchanged.}
\label{fig:RayDM_loop_diagrams}
\end{figure}

The Rayleigh operators are generated by attaching another external gauge-boson to the loop diagrams, as shown in Fig.~\ref{fig:RayDM_loop_diagrams}. In this case coupling to non-abelian gauge-bosons is possible as well. The Rayleigh scales associated with the non-abelian groups are
\be
\left\{\frac{1}{\LamR^3},\frac{1}{\LamRt^{3}}\right\} &=& \left(\frac{g^2\lambda^2C_f}{48 \mmedf^3 \pi ^2}\right) \left\{\mathcal{F},  \mathcal{\tilde{F}}\right\} 
\ee
up to corrections of order $\mathcal{O}\left(\mX/\mmeds \right)$. (See \cite{Ullio:1997ke} for the equivalent calculation for a neutralino.)
Here $C_f$ is defined through the generators of the representation of the matter in the loop tr$(t^a t^b) = C_f \delta^{ab}$ and is equal to $1/2$ for matter in the fundamental representation. In the abelian case this factor should be replaced by the square of the hypercharge times the size of the representation (so a factor of $2\times(1/2)^2 = 1/2$ for the matter we consider). The functions $\mathcal{F}(r)$ and $\mathcal{\tilde{F}} $ are given explicitly in the appendix. In general these are functions of all the masses in the problem as well as the Mandelstam variables $s,t$, and $u$.  To lowest order in an expansion in the messenger mass they are given by
\be
\mathcal{F}(r) &=& \frac{r^2 \left(2+ 3r^2-6 r^4+r^6+12 r^2 \log(r)\right)}{ \left(1-r^2\right)^3}, \\
\mathcal{\tilde{F}}(r) &=& \frac{r^2 \left(3-3r^4+12 r^2 \log(r)\right)}{ \left(1-r^2\right)^3}.
\ee
with $\mathcal{F}(1)=0$ and $\mathcal{F}(1)=1$. When evaluating the resulting cross-sections we use the full contribution to 1-loop order as discussed below.

When a pseudo-Dirac state is present at freezeout the annihilation rate is dominated by the s-channel annihilation into SM charged fermion pairs through a $\gamma/Z$ exchange. In the non-relativist limit appropriate for freezeout this is given by,
\be
\nonumber
\sigma &v&\left(\bar{\chi}\chi \rightarrow f\bar{f}\right) = \frac{\pi \alpha^2 ~q_f^2}{\mX^2} \frac{(c_{_{\rm W}}^{-1} F_1 +2 \mX \left(\frac{\mu_\chi}{e} \right)F_2 )^2}{\left(1-\mZ^2/4\mX^2\right)^2  } \\
&\times&\left( a_f^2 \sW^2+ \left( v_f  \sW+  \left(1-\mZ^2/4\mX^2\right) \cW \right)^2\right)
\ee
Here $\cW$ ($\sW$) is the cosine (sine) of the Weinberg angle, $q_f$ is the electric charge of the fermion and $v_f$ ($a_f$) is the ratio of its vector (axial) coupling to the $\ZZ$ boson to its electric coupling.

\begin{figure}
\begin{center}
\includegraphics[width=0.5 \textwidth]{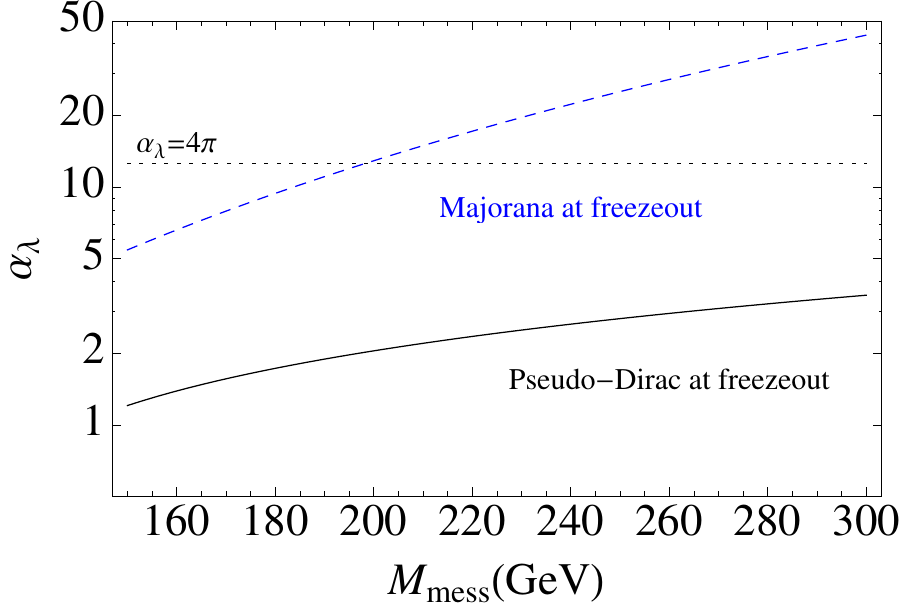}
\end{center}
\caption{The Yukawa coupling vs. the messengers' mass fixing the freezeout rate at $\sigma v = 3\times 10^{-26}{\rm cm^3/s}$ (solid) and $\sigma v = 6\times 10^{-26}{\rm cm^3/s}$ (dashed). The upper (blue) curves are in the case when a single Majorana state is present at freezeout. The lower (black) curves are for the case when a pseudo-dirac state is present where the annihilation into SM fermions dominates.  }
\label{fig:alphaVsMmed}
\end{figure}

When only a single Majorana state is present the annihilation rate is dominated by the RayDM operators which contribute through WIMP annihilation into $\gamma\gamma$,$\gamma \ZZ$, $\ZZ \ZZ$ and $\Wpm\Wmp$~\cite{Weiner:2012cb}. By fixing the annihilation rate through these operators we can obtain a simple relation between the messenger mass scale and the Yukawa coupling $\lambda$ in the case of a WIMP with mass at $130\GeV$,
\be
M_{\rm med} = 150\GeV ~\left( \frac{\alpha_\lambda}{5.5}\right)^{1/3} \left(\frac{\sigma v}{3\times 10^{-26}{\rm ~cm^3 s^{-1} }} \right)^{1/6}
\ee
The relation between the Yukawa coupling and the messenger mass in both scenarios is shown in Fig.~\ref{fig:alphaVsMmed}.  

\section{Scenarios for the Fermi Line}
With this in hand, we can explore the degree to which this UV scenario will actually provide the appropriate physics for the putative line signal observed by the Fermi satellite. We have essentially two scenarios to consider: MiDM, with a pseudo-Dirac state at freezeout and one with a pure Majorana state (where the heavier dark state is decoupled).

To begin with, it is worth emphasizing that except for the most strongly coupled region of parameter space for the MiDM scenario, the $\gamma \gamma$ signal in the present day is dominated by the dimension-7 Rayleigh operators, as argued in \cite{Weiner:2012cb}, where the hybrid scenario was proposed. Consequently, for both MiDM and RayDM scenarios, the ratios of $\gamma \gamma$ and $\gamma Z$ are determined essentially by the representation of the messengers. For doublet messengers, the relative sizes of the Rayleigh operators is $\cos \theta_\chi = g_Y^2/\sqrt{g_Y^4 + g^4} \approx 0.29$ (i.e., the sizes are determined by the gauge couplings alone). 

For a doublet, this yields a ratio of $\sigma_{\gamma\gamma}/(\sigma_{\gamma Z}/2) \simeq 2.2$ and $\sigma_{tot}/(\sigma_{\gamma \gamma}+1/2 \sigma_{\gamma Z}) \simeq 5$. Thus, it is a natural expectation for models dominated by Rayleigh annihilation from a doublet loop that both lines should be visible. Secondly, we emphasize that {\em there is no issue with constraints from continuum emission in these models.} The ratio of monochromatic to continuum photons is safely below the limits of \cite{Buchmuller:2012rc,Cohen:2012me,Cholis:2012fb}. This is true irrespective of whether the freezeout occurs through the dipole or through the Rayleigh operator (as the continuum ratio is the same in both cases). Essentially, continuum emission is a problem for models where the line annihilation is produced at a higher order in some perturbative expansion from the continuum. Here, they are produced at the same order and the lack of any a priori problem is obvious.

Let us focus for a moment at the MiDM scenario. Here, the freezeout occurs via the dipole annihilation into $f \bar f$, while the present day annihilation is dominated by the Rayleigh operator. For messengers in the 150-200 GeV range, the coupling $\alpha \approx 1$ is strong, but still perturbative, and the theory is calculable. The annihilation into photons $\vev{\sigma v}_{\gamma \gamma} + \tfrac{1}{2}\vev{\sigma v}_{\gamma Z} \approx 10^{-27}{\rm cm^3 s^{-1}}$ is precisely the right value to explain the observed excess. Intriguingly, the dipole here is slightly smaller than $10^{-3} \mu_N$, which is the size previously argued to explain the DAMA annual modulation result \cite{Chang:2010en}. Thus, it is conceivable (and is already strongly constrained from direct detection experiments) that such a scenario could also yield an explanation of the DAMA result.

In the case where we have only a single Majorana fermion (both in the present universe as well as at freezeout), we must have a truly strongly coupled theory to generate the Rayleigh operator of the appropriate size. Again, we have both $B_{\mu\nu} B^{\mu\nu}$ and $W^{\mu\nu}W_{\mu\nu}$ operators with $\cos \theta_\chi = 0.29$. Assuming that together, these provide the appropriate relic abundance we have a Rayleigh scale of $\sim 550 \gev$ (a difference of $2^{1/6}$ from the Dirac case). Here, normalizing to freezeout, we expect a cross section of $\sigma_{\gamma \gamma}+1/2 \sigma_{\gamma Z} \sim 6 \times 10^{-27} {\rm cm^3 s^{-1}}$, which is large, but perfectly acceptable for a slightly flatter halo.

\begin{figure}
\begin{center}
\includegraphics[width=0.5 \textwidth]{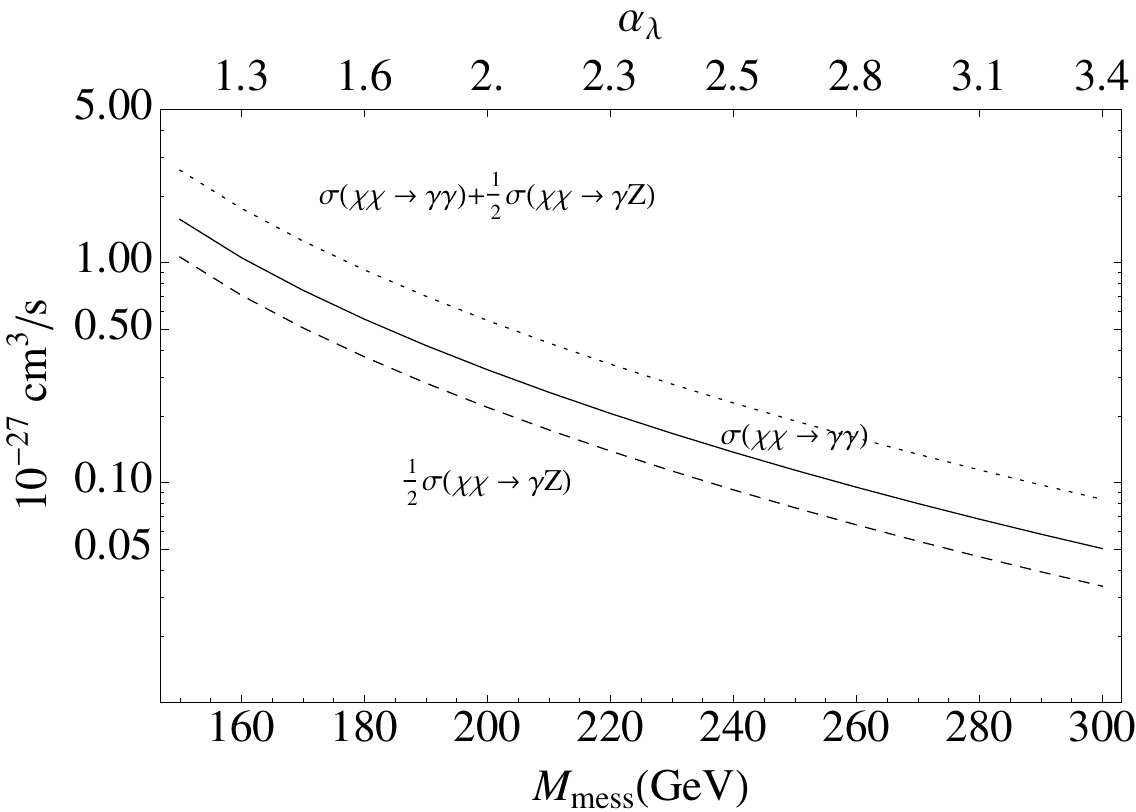}
\end{center}
\caption{The solid (dashed) curve depicts the annihilation rate $\bar{\chi}\chi\rightarrow \gamma\gamma$ ($\rightarrow \gamma\ZZ$) as a function of the mass of the messenger in the case of a pseudo-Dirac WIMP. The Yukawa coupling is fixed by requiring thermal freezeout with an annihilation cross-section of $\sigma v = 6\times 10^{-26}{\rm cm^3/s}$. }
\label{fig:ggRateVsMmed}
\end{figure}

\section{Conclusions}
The recent evidence for a possible dark matter signal in gamma rays has prompted a reexamination of the interactions of dark matter with light. For a Majorana fermion, the leading operators are a dimension-5 dipole operator in the presence of a nearby excited state (the MiDM scenario) or a dimension-7 Rayleigh operator more generally. The scales of the operators ($\sim \tev$ for the dipole and $\sim 600 \gev$ for the Rayleigh operator) suggest that the UV completion is at or near the weak scale.

In the presence of the simplest possible theory that generates these - namely, a loop of electroweakly charged messengers - we can understand the overall phenomenology for the 130 GeV line and more generally. We have found that for most of the parameter space, excepting only the most strongly coupled, the Rayleigh operator dominates the present-day $\gamma \gamma$ signals, while the dipole annihilation $\chi \chi^* \rightarrow f \bar f$ (which is inaccessible today) dominates freezeout. With weak-scale messenger masses and a strong, but perturbative ($\alpha \sim 1$) coupling, MiDM provides a natural framework to explain these signals, without any apparent conflict from the data. If the charged matter carries transforms under some strong gauge group such as in e.g., Sister Higgs models \cite{Alves:2012fx}, such couplings are not unreasonable. Intriguingly, the generated dipole is also of the size necessary to explain the DAMA modulation. Without the excited WIMP state for freezeout, annihilation through the Rayleigh operator also provides a viable scenario both for the 130 GeV line and relic abundance, but at the cost of both relatively light matter and very strong (non-perturbative) couplings.

In both scenarios the strength of $\sigma_{\gamma\gamma}/(\tfrac{1}{2}\sigma_{\gamma Z})$ is determined entirely (at leading order in weak couplings) by the $SU(2)\times U(1)$ representations of the matter in the loop. For $({\bf 2}, {\bf \pm 1/2})$ messengers, the ratio is roughly 2.2:1, consistent with recent suggestions.

The low scales of the new matter imply that the effective operator approach is limited in its quantitative applications. Indeed, including all orders in the $m_\chi/M_{mess}$ expansion tend to enhance the annihilation rates both in the late and early universe. Nonetheless, normalizing to the appropriate relic abundance, the present day signals are not dramatically changed when including these effects, only their interpretation in terms of the masses and couplings of the new states.

In summary, one can understand the effective theory of MiDM and RayDM with a simple UV completion that gives the relative signals and scales in different regions of parameter space. Remarkably, this simple completion is better than just a toy model, providing a successful theory at perturbative coupling. One must accept a somewhat strongly interacting theory, but there is no reason to think our calculational challenges prohibit nature from realizing such scenarios. Should the Fermi line prove to be true evidence of dark matter, this model may help provide qualitative and quantitative understanding of the signal. 

\begin{acknowledgments}
NW is supported by NSF grant \#0947827. IY is supported in part by funds from the Natural Sciences and Engineering Research Council (NSERC) of Canada. Research at the Perimeter Institute is supported in part by the Government of Canada through Industry Canada, and by the Province of Ontario through the Ministry of Research and Information (MRI). 
\end{acknowledgments}
\onecolumngrid
\begin{appendix}
\section{Calculating the Dipole and Rayleigh Coefficients and Form Factors}
\label{app:formulas}

In this supplement we provide explicit and detailed expression for the results quoted and used in the paper. We also offer simplified expression in some limits of physical interest. 


\subsection{Single gauge-boson vertex}
The amplitude with the gauge-boson attached to the scalar is 
\be
\label{eqn:MDM_loop_scalar}
&\quad& \nonumber \\
i\mathcal{M}_s~~ &=& \parbox[t]{5cm}{\vspace{-1cm}
  \includegraphics[scale=0.6]{MDM_loop_scalar.pdf} }  \nonumber \\ \nonumber 
\\\nonumber &=&  \bar{v}(p_2)\left( i\lambda\right) \Big( \loopint{4}{k} \frac{i\left(\slashed{p}_1-\slashed{k}+\mmedf \right)}{(p_1-k)^2 - \mmedf^2}   \frac{i}{k^2-\mmeds^2}  \left(-ig (2k-q)^\mu\right) \frac{i}{(k-q)^2-\mmeds^2} \Big) \left(- i\lambda\right) u(p_1)
\\\nonumber
\\\nonumber &=& 2\lambda^2 \bar{v}(p_2) \int d^3 x \Big( \frac{i}{2}\gamma^\mu~ \frac{\Gamma\left(2-\tfrac{d}{2}\right)}{(4\pi)^{d/2}}\frac{1}{\Delta_s^{2-d/2}} \\ 
&\quad&\quad\quad\hspace{20mm} + \frac{i}{32\pi^2} \frac{2\mX z\left((1-z)\mX + \mmedf \right)\left(\gamma^\mu + i\sigma^{\mu\nu}q_\nu/2\mX\right)}{\Delta_s}\Big)u(p_1).
\ee
Here $d^3x\equiv dx~dy~dz $ are the usual Feynman parameters, restricted to $x+y+z=1$ and we used dimensional regularization with $\epsilon=4-d$ to define the integral. The denominator is given by
\be
\Delta_s = z \mmedf^2 + (x+y) \mmeds^2 -z(1-z)\mX^2 -x y q^2
\ee
The divergent piece in Eq.~(\ref{eqn:MDM_loop_scalar}) cancels against the divergent piece in the second diagram where the gauge-boson is attached to the fermion line. The amplitude for this process is
\be
\label{eqn:MDM_loop_fermion}
&\quad& \nonumber \\
i\mathcal{M}_f~~ &=& \parbox[t]{5cm}{\vspace{-1cm}
  \includegraphics[scale=0.6]{MDM_loop_fermion.pdf} }  \nonumber \\ \nonumber 
&=&  \bar{v}(p_2)\left( i\lambda\right) \Big( \loopint{4}{k} \frac{i\left(-\slashed{p}_2-\slashed{k}+\mmedf \right)}{(-p_2-k)^2 - \mmedf^2} \left(i g \gamma^\mu \right) \frac{i\left(\slashed{p}_1-\slashed{k}+\mmedf \right)}{(p_1-k)^2 - \mmedf^2}  \frac{i}{k^2-\mmeds^2}  \Big) \left(- i\lambda\right) u(p_1) \\ \nonumber
&=& 2\lambda^2 \bar{v}(p_2) \int d^3 x \Big( \frac{i}{2}\gamma^\mu~\left(1-\tfrac{d}{2}\right) \frac{\Gamma\left(2-\tfrac{d}{2}\right)}{(4\pi)^{d/2}}\frac{1}{\Delta_f^{2-d/2}} \\
&\quad&\quad\quad\hspace{20mm}  -\frac{i}{32\pi^2}\frac{\big( (z\mX+\mmedf)^2 + x y~ q^2\big)\gamma^\mu - 2i~ x \left(z\mX+\mmedf\right) \sigma^{\mu\nu}q_\nu }{\Delta_f}
 \Big)u(p_1).
\ee
with
\be
\Delta_f = z \mmeds^2 + (x+y) \mmedf^2 -z(1-z)\mX^2 -x y ~q^2
\ee
combining both diagrams the limit $\epsilon\rightarrow 0$ yields a finite expression. The result can be written in terms of two form-factors as in Eq.~(3) in the text
\be
\label{eqn:app:GeneralVertex}
\Gamma^\mu(q^2) = \gamma^\mu F_1(q^2) + i \left( \frac{\mu_\chi}{2}\right) \sigma^{\mu\nu}q_\nu F_2(q^2)
\ee
The dipole strength $\mu_\chi$ is given by
\be
\mu_\chi = \frac{g'\lambda^2}{32\pi^2\mmedf},
\ee
and the form-factors are 
\be
F_1(q^2;\mX,\mmedf,\mmeds) = \frac{g' \lambda^2}{16\pi^2}\int d^3x ~\left( 1 + \frac{2\mX\left(\mmedf+(1-z)\mX \right)}{\Delta_s} - \frac{\left(\mmedf+z\mX \right)^2+x y~q^2}{\Delta_f} + \log\left( \frac{\Delta_f}{\Delta_s} \right) \right)
\ee
and
\be
F_2(q^2;\mX,\mmedf,\mmeds) = 2 \mmedf \int d^3x ~\left(\frac{z (\mmedf + (1-z)\mX)}{\Delta_s} + \frac{2 x (\mmedf+ z \mX)}{\Delta_f}   \right)
\ee
When the momentum exchange is small $q^2\rightarrow 0$ the charge form-factor approaches zero as it should since the WIMP state is uncharged
\be
F_1(q^2) \xrightarrow{q^2\rightarrow 0} -\frac{ \mu_\chi q^2}{6 \mmedf} \frac{2r^2 \left(3 r^2-3-\left(2+r^2\right) \log r^2\right)}{ \left(1-r^2\right)^2} +\mathcal{O}\left( \mmedf^{-3}\right)
\ee
where $r = \mmedf/\mmeds$. Similarly, the dipole form-factor can be simplified in the limit of heavy messengers
\be
F_2(q^2) \xrightarrow{\mmedf,\mmeds\gg q^2,\mX} \frac{2 r^2 \left(r^2-1-\log r^2\right)}{\left(1-r^2\right)^2} +\mathcal{O}\left(\mmedf^{-1} \right).
\ee
More importantly are the simplification that occur in the non-relativistic limit relevant in the case of the annihilation into SM fermion pairs, $\bar{\chi}\chi\rightarrow \bar{f} f$ where $q^2 = 4\mX^2$ to lowest order in velocity. For simplicity we only consider the case of equal messenger masses since the most general case yields extremely complex and unilluminating formulas. The charge form-factor is
\be
F_1\Big|_{q^2=4\mX^2} = -\frac{g' \lambda^2 }{60 \pi^2 \tilde{m}_\chi^5 \sqrt{1-\tilde{m}_\chi^2}} &\Big(& 2 \tilde{m}_\chi^3 \left(1-\tilde{m}_\chi^2\right)^{3/2}+\left(-16+18 \tilde{m}_\chi^2-3 \tilde{m}_\chi^4+\tilde{m}_\chi^6\right) \arcsin\left( \tilde{m}_\chi\right) 
\\\nonumber
&+&\left(4-\tilde{m}_\chi^2\right)^2 \sqrt{4-5 \tilde{m}_\chi^2+\tilde{m}_\chi^4} \arctan\left(\frac{\tilde{m}_\chi}{\sqrt{4-\tilde{m}_\chi^2}}\right)\Big)
\\
F_2\Big|_{q^2=4\mX^2} = -\frac{2}{15 \tilde{m}_\chi^6 \sqrt{4-\tilde{m}_\chi^2}} &\Big(&2 \sqrt{1-\tilde{m}_\chi ^2} \sqrt{4-\tilde{m}_\chi ^2} \left(-16+\tilde{m}_\chi ^2 (12+\tilde{m}_\chi  (15+4 \tilde{m}_\chi ))\right)\arcsin\left(\tilde{m}_\chi \right)
\\\nonumber
&~& \hspace{-40mm}+\big(\tilde{m}_\chi ^3 \sqrt{4-\tilde{m}_\chi ^2} \left(4+\tilde{m}_\chi ^2\right)+2 (-2+\tilde{m}_\chi ) (2+\tilde{m}_\chi )^2 (-8+\tilde{m}_\chi  (4+\tilde{m}_\chi  (7+4 \tilde{m}_\chi ))) \arctan\big(\frac{\tilde{m}_\chi }{\sqrt{4-\tilde{m}_\chi ^2}}\big)\big)\Big)
\ee
where $\tilde{m}_\chi$ is the mass of the WIMP in units of the messenger mass. Since the WIMP is light than the messengers a power expansion in $\tilde{m}_\chi$ is likely more useful and illuminating
\be
F_1\Big|_{q^2=4\mX^2} &=& -\left(\frac{g'\lambda^2}{ 48\pi^2} \right) \left(\tilde{m}_\chi^2+\frac{19 \tilde{m}_\chi^4}{30}+\frac{19 \tilde{m}_\chi^6}{42}+\frac{29 \tilde{m}_\chi^8}{84} \right) + \mathcal{O}\left(\tilde{m}_\chi^{-10} \right)\\
F_2\Big|_{q^2=4\mX^2} &=& 1+\frac{\tilde{m}_\chi}{3}+\frac{\tilde{m}_\chi^2}{2}+\frac{7 \tilde{m}_\chi^3}{45}+\frac{3 \tilde{m}_\chi^4}{10} +\mathcal{O}\left(\tilde{m}_\chi^{-5} \right)
\ee

\begin{figure}
\begin{center}
\includegraphics[width=0.6 \textwidth]{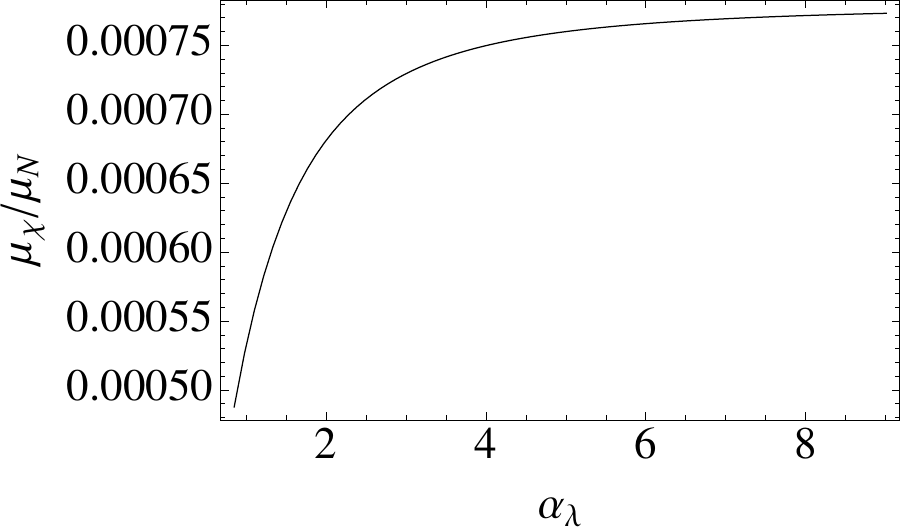}
\end{center}
\caption{The WIMP's dipole strength in units of the nuclear magneton $\muN$ as a function of its coupling to the heavier charged states. For a given choice of the coupling, the messenger mass is set by requiring the annihilation rate to SM fermion - anti-fermion pair to be equal to $3\times 10^{-26}{\rm cm^3/s}$. The curve asymptotes at large coupling since at that point the annihilation rate is dominated by the dipole contribution. }
\label{fig:MDM_from_loop}
\end{figure}

In Fig.~\ref{fig:MDM_from_loop} we plot the WIMP's dipole strength generated from the loop as a function of the coupling to the heavier charged states. Evidently, the coupling needs to be sizable in order to obtain a sufficiently large dipole even for charged states which are only slightly heavier than the WIMP. 

\subsection{Two gauge-boson vertex}
There are a total of seven loop diagrams contributing to the two gauge-boson vertex, three of which are related by a simple exchange of the two gauge-bosons. The first diagram where both external bosons are connected to the fermion line is the most complicated one and is given by
\be
\label{eqn:diagram1}
&\quad& \nonumber \\
i\mathcal{M}_1~~ &=& \parbox[t]{5cm}{\vspace{-1cm}
  \includegraphics[scale=0.6]{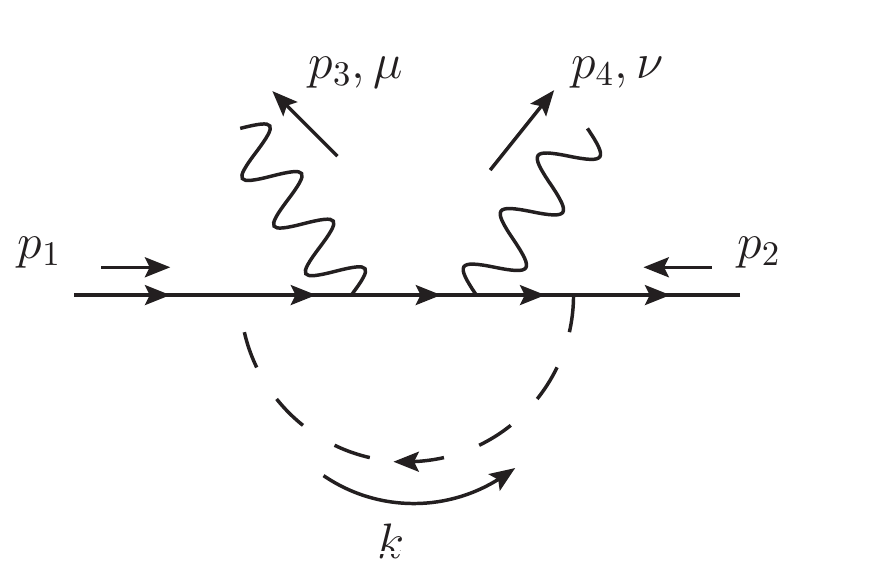} }  \nonumber \\  
&~& \hspace{-10mm} =  -3! g^2 \lambda^2 \bar{v}(p_2)\Bigg(\int d^4 x \loopint{4}{k} \frac{\left(-\slashed{k}-\slashed{p}_2+\mmedf \right) \gamma^\nu \left(\slashed{p}_1-\slashed{p}_3-\slashed{k}+\mmedf \right) \gamma^\mu \left(\slashed{p}_1-\slashed{k}+\mmedf \right)}{\left(k'^2 - \Delta_1\right)^4}   \Bigg) u(p_1)
\ee
where
\be
k &=& k' + \left(x_1 p_1 - x_2 p_2 +x_3(p_1 - p_3)\right) \\
\Delta_1 &=& \mmedf^2 (x_1 + x_2 + x_3)+ \mmeds^2 x_4 -s x_1 x_2 - t x_3 - m_3^2 x_1 x_3 + t x_1 x_3 + m_3^2 x_2 x_3\\ \nonumber  &-& s x_2 x_3 - 
 u x_2 x_3  + t x_3^2  + 
 \mX^2 (x_1^2 - x_1 (1 - 2 x_2 - x_3) - x_2 (1 - x_2 - 3 x_3)) 
\ee
Here $s, t, u$ are the usual Mandelstam variables ($s=(p_1+p_2)^2$, $t=(p_1-p_3)^2$, and $u=(p_1-p_4)^2$) and  $m_{3,4}^2 = p_{3,4}^2$ are the external bosons' mass, which we keep explicit as a check on the calculation since they should drop out when extracting the final answer for the coefficient of the Rayleigh terms. Shifting the momentum to $k'$, the momentum integral is finite and easily doable. The second amplitude, $i\mathcal{M}_2$, has the same form but with the two external bosons exchanged, 
\be
i\mathcal{M}_2 = i\mathcal{M}_1(\mu \leftrightarrow \nu, m_3 \leftrightarrow m_4, p_3 \leftrightarrow p_4, t \leftrightarrow u)
\ee
The next diagram has one external gauge-boson attached to the fermion line while the other to the scalar line
\be
\label{eqn:diagram3}
&\quad& \nonumber \\
i\mathcal{M}_3~~ &=& \parbox[t]{5cm}{\vspace{-1cm}
  \includegraphics[scale=0.6]{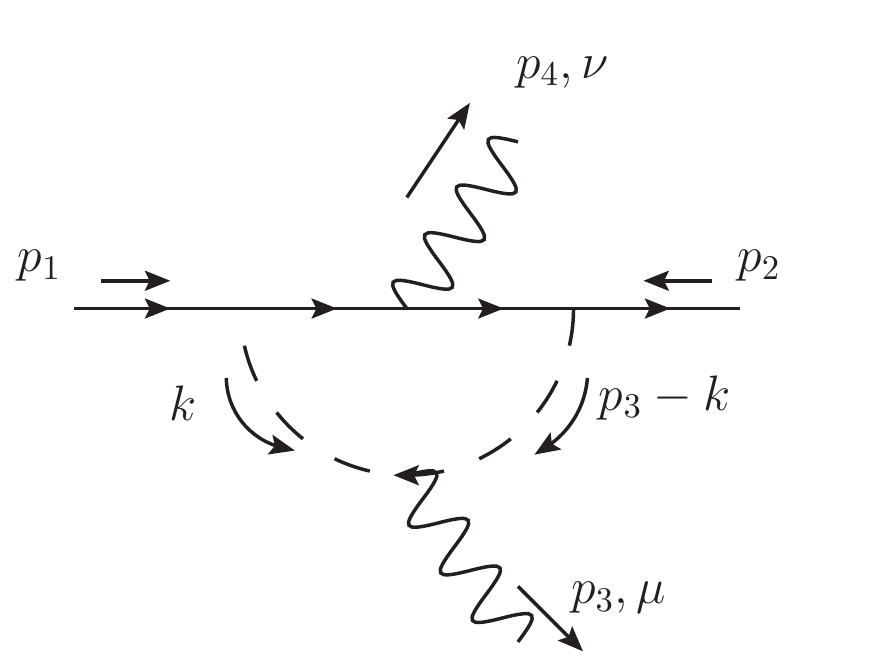} }  \nonumber \\ \nonumber 
\\
&=& 3! g^2 \lambda^2 \bar{v}(p_2)\Bigg( \int d^4 x \loopint{4}{k} \frac{\left(\slashed{p}_3-\slashed{p}_2-\slashed{k}+\mmedf \right) \gamma^\nu \left(\slashed{p}_1-\slashed{k}+\mmedf \right) \left(2k-p_3\right)^\mu}{\left(k'^2 - \Delta_3 \right)^4} \Bigg) u(p_1)
\ee
where,
\be
k &=& k' + \left(x_1 p_1 + x_2 (p_3-p_2) +x_3 p_3\right) \\
\Delta_3 &=& \mmedf^2 (x_1 + x_2) + \mmeds^2 (x_3+x_4) -u x_2 + m_3^2 x_1 x_2 - s x_1 x_2 - t x_1 x_2 + u x_2^2  - 
 m_3^2 x_3  \\\nonumber &+& m_3^2 x_1 x_3 - t x_1 x_3 + m_3^2 x_2 x_3 + u x_2 x_3 + 
 m_3^2 x_3^2 + \mX^2 (x_1^2 - x_2 x_3 - x_1 (1 - 3 x_2 - x_3)) 
 \ee
The fourth amplitude, $i\mathcal{M}_4$, is similar but with the two external gauge-bosons exchanged. The next diagram is
\be
\label{eqn:diagram5}
&\quad& \nonumber \\\nonumber
i\mathcal{M}_5~~ &=& \parbox[t]{5cm}{\vspace{-1cm}
  \includegraphics[scale=0.6]{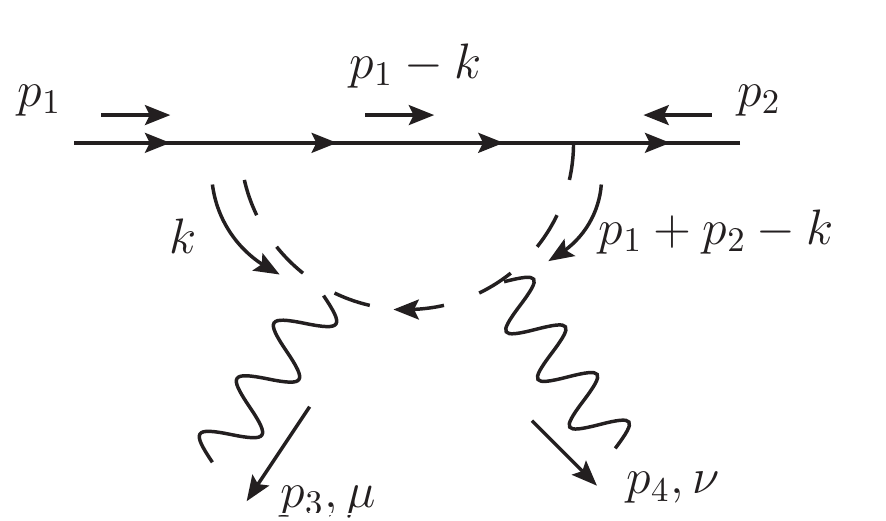} } 
\\ 
&=& -3! g^2 \lambda^2 \bar{v}(p_2)\Bigg( \int d^4 x \loopint{4}{k} \frac{\left(\slashed{p}_1-\slashed{k}+\mmedf \right) \left(2k-p_3\right)^\mu \left(2k-p_1-p_2-p_3\right)^\nu}{\left(k'^2 - \Delta_5 \right)^4}\Bigg) u(p_1)
\ee
where,
\be
k &=& k' +\left(x_1 p_1 + x_2 (p_1+p_2) +x_3 p_3\right) \\
\Delta_5 &=& \mmedf^2 x_1 + \mmeds^2(x_2+x_3+x_4) - s x_2 + s x_1 x_2 + s x_2^2 - m_3^2 x_3 + 
 m_3^2 x_1 x_3 - t x_1 x_3 \\\nonumber &+& 2 m_3^2 x_2 x_3 - t x_2 x_3 - u x_2 x_3 + m_3^2 x_3^2 +
  \mX^2 (x_1^2 - x_1 (1 - x_3) + 2 x_2 x_3) 
\ee
The sixth amplitude, $i\mathcal{M}_6$, is similar but with the two external gauge-bosons exchanged. The final diagram is
\be
\label{eqn:diagram7}
&\quad& \nonumber \\
i\mathcal{M}_7~~ &=& \parbox[t]{5cm}{\vspace{-1cm}
  \includegraphics[scale=0.6]{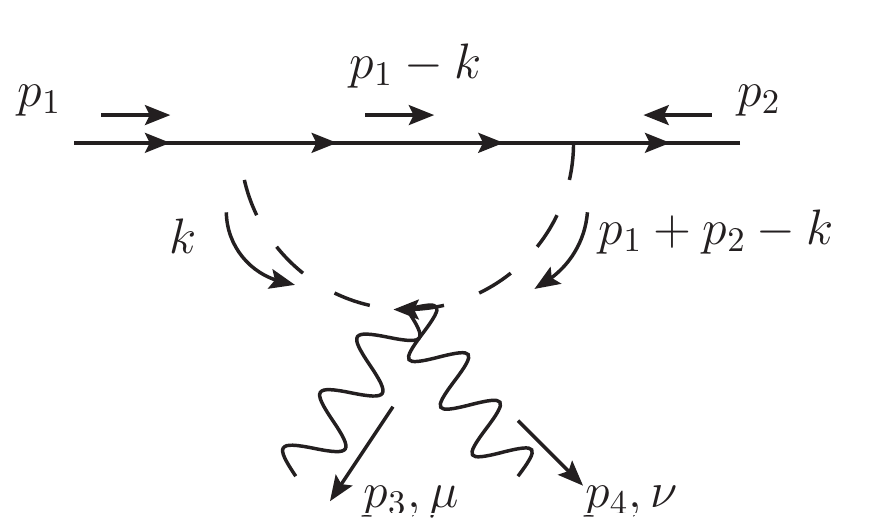} }  \nonumber \\  
&=& 2\times2! g^2\lambda^2 \bar{v}(p_2)\Bigg(  \int d^3 x \loopint{4}{k} \frac{\left(\slashed{p}_1-\slashed{k}+\mmedf \right)g^{\mu\nu}}{\left(k'^2 - \Delta_7 \right)^3}\Bigg) u(p_1)
\ee
where,
\be
k &=& k' + \left(x_1 p_1 + x_2 (p_1+p_2)\right) \\
\Delta_7 &=& \mmedf^2 x_1  + \mmeds^2 (x_2+x_3)- \mX^2 (1 - x_1) x_1 - s x_2 + s x_1 x_2 + s x_2^2 
\ee

Expanding the amplitudes in powers of inverse messenger mass the sum of the diagrams can be shown to vanish up to order $\mathcal{O}\left(M_{\rm mess}^{-3} \right)$. At this order the first contribution to the Rayleigh operators appear. The matching between the coefficient of the terms that appear in the Lagrangian and the terms in the amplitude is as follows,
\be
\nonumber
\bar{\chi}\chi  F_{\mu\nu}F^{\mu\nu} \quad &\iff &\quad 4 \bar{v}(p_2) u(p_1)~ \Big(p_3 \cdot p_4 g^{\mu\nu} -  p_3^\nu p_4^\mu\Big) \varepsilon_{\mu}(p_3)\varepsilon_{\nu}(p_4)
\ee
and
\be
\nonumber
i\bar{\chi}\gamma_5\chi  F_{\mu\nu}\tilde{F}^{\mu\nu} \quad &\iff& \quad  i\bar{v}(p_2)\Big(-\frac{i}{4!}\epsilon_{\rho\sigma\kappa\lambda} \gamma^{\rho}\gamma^{\sigma}\gamma^{\kappa}\gamma^{\lambda} \Big) u(p_1)\Big(4\epsilon^{\mu\alpha\nu\beta}p_{3\alpha}p_{4\beta}\Big)\varepsilon_{\mu}(p_3)\varepsilon_{\nu}(p_4)
\ee
Here $\varepsilon^{\mu}(p_3)$ and $\varepsilon^{\nu}(p_4)$ are the external gauge-boson's polarization vectors. In the large messenger mass expansion the Rayleigh scales are found to be
\be
\frac{1}{\LamR^3}&=& \left(\frac{g^2\lambda^2}{48 \mmedf^3 \pi ^2}\right)\mathcal{F}(r),\\
\frac{1}{\LamRt^3} &=& \left(\frac{g^2\lambda^2}{48 \mmedf^3 \pi ^2}\right)\mathcal{\tilde{F}}(r),
\ee
with
\be
\mathcal{F}(r) &=& \frac{r^2 \left(2+ 3r^2-6 r^4+r^6+12 r^2 \log(r)\right)}{ \left(1-r^2\right)^3}, \\
\mathcal{\tilde{F}}(r) &=& \frac{r^2 \left(3-3r^4+12 r^2 \log(r)\right)}{ \left(1-r^2\right)^3}.
\ee
Here $r =\mmedf/\mmeds$ is the ratio of the scalar messenger's mass to that of the fermion messenger. The form-factors obey $\mathcal{F}(1) =0$ and $\mathcal{\tilde{F}}(1) =1$. A plot of the relative strength of these two coefficients is shown in Fig.~(\ref{fig:FFvsFFdual}) below. 

\begin{figure}
\begin{center}
\includegraphics[width=0.45 \textwidth]{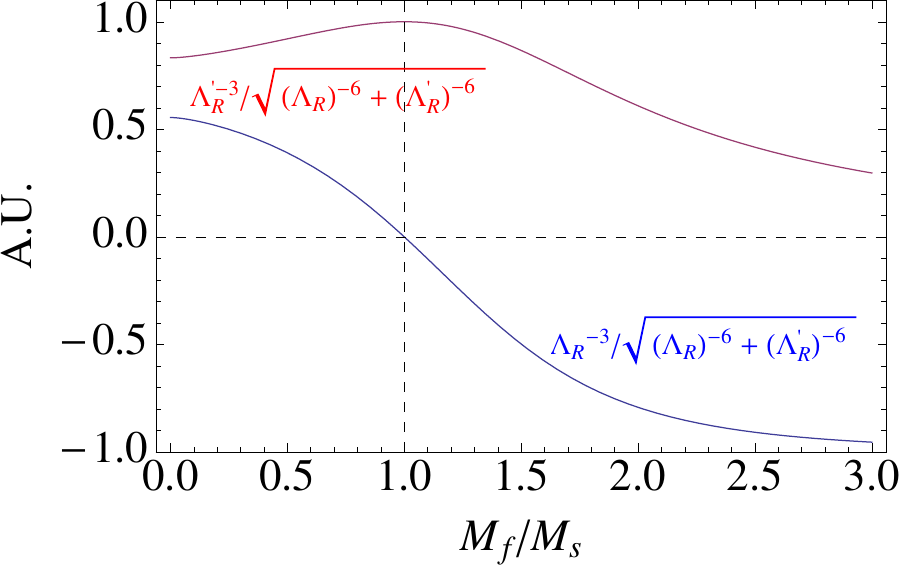}
\includegraphics[width=0.45 \textwidth]{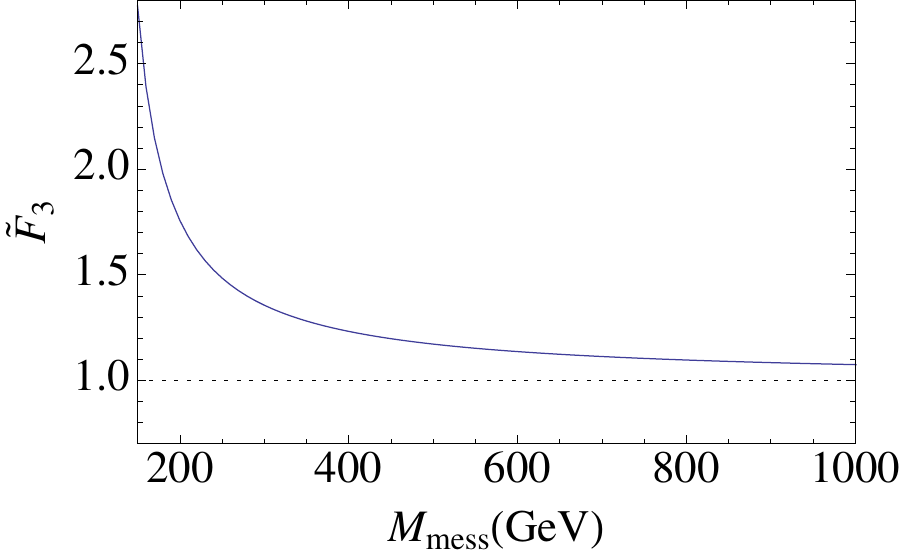}
\end{center}
\caption{On the left pane is a plot of the relative strength of the coefficients of the two Rayleigh operators as a function of the ratio of the fermion to scalar messenger. On the right pane is a plot of the $F_3'$ form-factor as a function of the messenger mass, assuming $\mX = 130\GeV$ and messengers of equal mass. $F_3'$ asymptotes to unity in the large messenger limit.}
\label{fig:FFvsFFdual}
\end{figure}

In the case of non-relativistic annihilation these results are inadequate when the messenger mass is not much larger than the WIMP mass because the kinematical variable $s=4\mX^2$ is of the same order (or larger) than the messenger mass. Not only is the expansion in inverse powers of the messengers mass is inadequate, one must worry about higher dimensional operators that could contribute. However, as we now argue, to one-loop order in perturbation theory such higher order operators do not contribute. In other words, no other terms in the amplitude aside from those associated with $i\bar{\chi}\gamma_5\chi  F_{\mu\nu}\tilde{F}^{\mu\nu}$ contribute to the non-relativistic cross-section. 

We first note that any terms with $p_{1,2}\cdot \varepsilon(p_{3,4})$ are velocity suppressed and can be neglected since the incoming momenta (which are mostly time-like) are contracted with the outgoing polarizations (which are space-like). Any terms with $p_3\cdot \varepsilon(p_{3})$ and $p_4\cdot \varepsilon(p_{4})$ vanish identically because of the transversality of the gauge-boson's polarizations. Therefore, by momentum conservation, $p_4\cdot \varepsilon(p_{3})$ and $p_3\cdot \varepsilon(p_{4})$ are both velocity suppressed and can be neglected. Thus, any term with external momenta contracted against the gauge-bosons' polarization can be neglected. 

Aside from the terms associated with the dual Raleigh operator, $i\bar{\chi}\gamma_5\chi  F_{\mu\nu}\tilde{F}^{\mu\nu}$, the only surviving terms are ones associated with $i\bar{\chi}\chi  F_{\mu\nu}F^{\mu\nu}$ which is itself velocity suppressed. So we need only extract the coefficient of the dual Raleigh operator albeit to all orders in the messengers mass. For simplicity we quote the result in the case of equal messenger mass
\be
\frac{1}{\LamRt^3} &=& \left(\frac{g^2\lambda^2}{48 \mmedf^3 \pi ^2}\right)\tilde{F}_3\left(s,t,u\right)
\ee
where the form-factor $\tilde{F}_3$ is given by the integral
\be
\tilde{F}_3\left(4\mX^2,-\mX^2,-\mX^2\right) = \int d^4x &\Bigg( &
\frac{3 \left(2 \tilde{m}_\chi+4 x_3 +\tilde{m}_\chi^2 \left(2 x_4+(1-x_3) (-3 x_4+2 (2 x_1+x_4) (2 x_2+x_4))\right)\right)}{\tilde{m}_\chi \left(1-\tilde{m}_\chi^2 (-x_4+(2 x_1+x_4) (2 x_2+x_4))\right)^2} 
\\
&-&\frac{3}{\tilde{m}_\chi \left(1+\tilde{m}_\chi^2 (x_1-x_2) (x_3-x_4)\right)}\Bigg)
\ee
The enhancement due to this form-factor is plotted on the right pane of Fig.~\ref{fig:FFvsFFdual}. 

\subsection{Cross-Sections}
The non-relativistic annihilation cross-sections associated with the two gauge-boson vertex (RayDM) were taken from ref.~\cite{Weiner:2012cb} and are reproduced here for completion. The general expression is given by 
\be
\sigma(\chi\chi\rightarrow VV)v &=& \frac{ g_{_{VV}}^2 }{4\pi}  \frac{\mX^4 }{\LamR^6}~\mathcal{K}_{_{VV}}
\ee
with the kinematic functions $\mathcal{K}_{_{VV}}$ and couplings $g_{_{VV}}$ defined as
\vspace{5mm}
\begin{eqnarray}
\label{eqn:gVV}
\mathcal{K}_{\gamma\gamma} = 1 \quad \quad \quad \quad\quad\quad &\quad& \quad g_{\gamma\gamma} = \cos\thetaX \cos^2\theta_W + \sin\thetaX \sin^2\theta_W \\
\mathcal{K}_{\rm \gamma {\scriptstyle Z}} = 2\left(1-\frac{\mZ^2}{4\mX^2} \right)^{3}  \quad &\quad&  \quad g_{\rm \gamma {\scriptstyle Z}} = \tfrac{1}{2}\left(\sin\thetaX- \cos\thetaX\right) \sin(2\theta_W) \\ 
\mathcal{K}_{_{\rm ZZ}} =  \left(1-\frac{\mZ^2}{\mX^2}\right)^{3/2} \quad &\quad& \quad g_{_{\rm ZZ}} = \cos\thetaX \sin^2\theta_W + \sin\thetaX \cos^2\theta_W \\
\mathcal{K}_{_{WW}} =  2 \left(1-\frac{\mW^2}{\mX^2}\right)^{3/2} \quad &\quad& \quad g_{_{WW}} =  \sin\thetaX
\end{eqnarray}
In the case of messengers in the $({\bf 2,\tfrac{1}{2}})$ representation of $\SUWeak$ the angle $\cos\thetaX = g'^2/\sqrt{g'^4 + g^4} =0.29$. 

The most important annihilation cross-section associated with the single gauge-boson vertex is that of the WIMPs into SM fermion pairs. If present, this annihilation mode dominates over all other channels. Using the general expression for the one gauge-boson vertex, Eq.~(\ref{eqn:app:GeneralVertex}) we obtain the following matrix element
\be
\frac{1}{4}\sum_{\rm pol}\left|\mathcal{M}\right|^2 &=& \frac{e^4q_f^2}{\cW^2\left(s-\mZ^2\right)^2 s^2} \Bigg( F_1^2\left(2 \mX^4 + s^2 - 4 \mX^2 t + 2 s t + 2 t^2 \right) + F_1 F_2\left(4 \mX s^2 \left(\frac{\mu_\chi}{e}\right)\cW \right) 
\\\nonumber
&+& F_2^2\left(-s (\mX^4 - 2 \mX^2 (s + t) + t (s + t)) \left(\frac{\mu_\chi}{e}\right)^2 + 
 s (-\mX^4 + 2 \mX^2 (s + t) - t (s + t)) \left(\frac{\mu_\chi}{e}\right)^2 \left(\cW^2 - \sW^2 \right)\right) \Bigg)
\\\nonumber
&\times& \Big(\mZ^4-2 \mZ^2 s+s^2 \left(1+a_f^2+v_f^2\right)+\left(\mZ^4-2 \mZ^2 s+s^2 \left(1-a_f^2-v_f^2\right)\right) \left(\cW^2-\sW^2\right)+4 s \left(s-\mZ^2\right) v_f \sW\cW \Big)
\ee
where $s$ and $t$ are the usual Mandelstam variables $s=(p_1+p_2)^2$ and $t=(p_1-p_3)^2$. 

\end{appendix}
\onecolumngrid

\bibliography{RayDMbib}

\end{document}